\documentclass{aa501}

\usepackage{graphicx}
\begin{document}
   \title{NICS-TNG infrared spectroscopy of trans-neptunian objects 2000 EB173 and 2000
   WR106}

   \author{J. Licandro
          \inst{1}
          \and
          E. Oliva
	  \inst{1,2}
	  \and
	  M. Di Martino
	  \inst{3}
%	  \fnmsep
          }

   \offprints{J. Licandro, e-mail licandro@tng.iac.es}

   \institute{Centro Galileo Galilei \& Telescopio Nazionale Galileo, P.O.Box 565, E-38700, S/C de La Palma, Tenerife,
    		Spain.\\
         	\and
	     Osservatorio di Arcetri, Largo E. Fermi 5, I-50125 Firenze,
	     Italy.\\
	     \and
             Osservatorio Astronomico di Torino, I-10025 Pino Torinese, Italy.\\
             }

  \date{Received April 12, 2001; accepted May 11, 2001}

   \abstract{
   We report complete near-infrared (0.9-2.4 $\mu$m) spectral observations of trans-neptunian objects (TNOs) 
   2000 EB173 and
   2000 WR106 collected using the new Near Infrared Camera Spectrometer (NICS) attached
   to the 3.56m Telescopio Nazionale Galileo (TNG). 
   Both spectra are very red and with a quite strong and broad drop extending throughout
   the K band. However, while 2000 EB173 does not show any evidence of narrow absorption
   features, the spectrum of 2000 WR106 has quite deep water ice absorption at 1.5 and
   2.0 $\mu$m. Moreover, the latter object is significantly less red than the former indicating,
   therefore, that the surface of 2000 WR106 is ``cleaner'' (i.e. less processed by particle
   irradiation) than that of 2000 EB173. 
   %The detection of water ice in 2000 WR106 also suggests
   %that this object could have a geometric albedo higher than the 4\% typically assumed.
   }
      
   \titlerunning{NICS-TNG infrared spectroscopy of 2000 EB173 and 2000 WR106}
   
   \maketitle
%
%________________________________________________________________

\section{Introduction}

   The population of objects in the region just beyond the orbit of Neptune (TNOs),
   called  the Edgeworth-Kuiper belt (EKb),
   are remnant planetesimals from the early solar system formation stages 
   (Edgeworth \cite{Edgeworth49}; Kuiper \cite{Kuiper51}).
   They probably contain some of the least modified materials remaining from the protosolar
   nebula, though the study of their surface properties is very important from a
   cosmogonical point of view. Their study could provide important clues to understand the
   conditions 
   existing at the beginning of the solar system.
   
   Near-infrared spectroscopy is a powerful mean for remote determination of the composition of
   volatile surface component of the outer solar system objects (Brown \& Cruikshank \cite{BrownCrui97}), but,   
   due to the small size and distance of this objects, this technique is actually limited to
   the brighter members of the EKb. In fact, only 4 TNOs have published
   near-infrared spectra, and all obtained with the Keck 10m telescope. Brown et al. (\cite{Brownetal97})
   present the spectra of 1993 SC; Luu \& Jewitt (\cite{LuuJew98}) the spectra of 1996 TL66;
   Brown et al. (\cite{BrownCrPe99}) the one of 1996 TO66; and finally  Brown et al. (\cite{Brownetal2000}) 
   the spectra of 
   2000 EB173. Among them, only for 1996 TL66 the spectrum spans the whole near-infrared
   range, while for the others the spectra cover the H and K bands (1.4 - 2.4 $\mu$m) only. 
   Even though
   there are very few TNO spectra published, different surface properties have been inferred
   among them. The spectra of 1993 SC show absorption bands possibly due to hydrocarbons, the one of 
   1996 TO66 shows absorptions due to water ice, while the spectra of 1996 TL66 and 
   2000 EB173 are featureless in the observed range. This diversity of surface 
   compositions has been observed also
   in similar objects like Centaurs and irregular satellites (Brown \cite{Brown2000}).
   
   During the commissioning phase of NICS at TNG, we had the opportunity to observe two of the
   brighter known TNOs: 2000 EB173 and 2000 WR106.(V=19.7 and 20.1, respectively). Assuming a
   geometric albedo of 4\%, these objects have diameters of 670 and 1100 km respectively . 
   If the typically assumed albedo is
   correct, 2000 WR106 is the larger known TNO after Pluto. 
   From the dynamical point of view, the 2000 EB173 orbital elements suggest that the orbit
   is compatible with a Pluto-like 2:3 resonance, and as Pluto, it is a Neptune--crosser. Its
   orbital elements are very similar to those of TNO 1995 QY9. On the other hand, 2000 WR106 has orbital
   elements that fit with those of the classical TNO group. 
 
%__________________________________________________________________

\section{Observations}

   We have obtained low resolution spectra of 2000 WR106 and 2000 EB173 on January 14 and
   February 5, 2001 respectively, with the 3.6m Telescopio Nazionale Galileo (TNG) 
   during the commissioning phase of NICS,
   the near-infrared camera and spectrometer expressly designed and built for the TNG. NICS is
   a FOSC-type cryogenic focal reducer, equipped with two interchangeable cameras feeding a
   Rockwell Hawaii 1024x1024 array. The camera used for the spectroscopic observations
   has a focal ratio of F/4.3, corresponding to a plate scale = 0.25"/pixel 
   (Oliva \& Gennari \cite{OlivaGen95}; Baffa et al. \cite{Baffaetal00}). 
   Among the many imaging and spectroscopic observing modes, NICS offers a unique, high throughput,
   low resolution spectroscopic mode with an Amici prism disperser (Oliva \cite{oliva01}),
   which yields a complete 0.9-2.5 $\mu$m spectrum. A 1.0" width slit corresponding
   to a spectral resolving power $R\simeq50$ and quasi-constant along the spectrum, has
   been used.
   The  low resolution together with the high efficiency of the Amici prism (about 90\%
   in all the infrared range) allowed us to obtain spectra of faint objects like TNOs with a
   four meter class telescope for the first time, and 
   with the advantage of having the whole infrared range measured simultaneously. 
   
   The identification of the TNOs was done by taking images through the J$_s$ filter the day
   of the observations and the day before, and by comparing them. The objects were identified as 
   moving objects at the predicted position and with the predicted proper motion.
   The slit was oriented in the direction of the motion of the TNOs,  
   so they remained in the slit during the integration tracking at the sidereal rate. 
   The acquisition
   consisted of a series of 3 images of 90 seconds exposure time in one position (position {\em A}) of 
   the slit for 2000 EB173 (4x60 s for 2000 WR106) and then offsetting the telescope by 40" in the
   direction of the slit (position {\em B}). This process 
   was repeated and a number of {\em ABBA} cycles were acquired. The total exposure time
   was 3360 seconds for 2000 WR106, and 4320 seconds for 2000 EB173 respectively.
   The reduction of the spectra was done by subtracting consecutive {\em A} and {\em B}
   images. 
   Each {\em A-B} frame had some residual of sky emission, related to sky transparency fluctuations
   and/or intrinsic variation of the air-glow atmospheric emission, which was eliminated by
   extracting the ``short-slit'' spectrum around each $A$, $B$ spectrum, then aligning and combining
   them into the final short-slit spectrum from which the 1D spectra were extracted. No
   contribution from residual sky emission was found within the noise of the spectra. 
%   In this way the sky contribution is subtracted,
%   resulting two spectra on each image, one positive and other negative. Then shifting the 
%   {\em B - A} image in such way that both positive spectra coincides, and coading {\em A - B}
%   and the shifted {\em B - A} images, the resulting spectrum have the signal of the {\em A} and {\em
%   B} ones, and any residuals of the sky subtraction due to sky variation during the time span
%   between images dissapear. All the so obtained spectra were averaged to obtain the final spectrum
%   of the TNOs. 
   Wavelength calibration was performed using an Argon lamp and the deep telluric 
   absorption features. 
   
\begin{figure*}
   \centering
   \includegraphics[angle=-90,width=\textwidth]{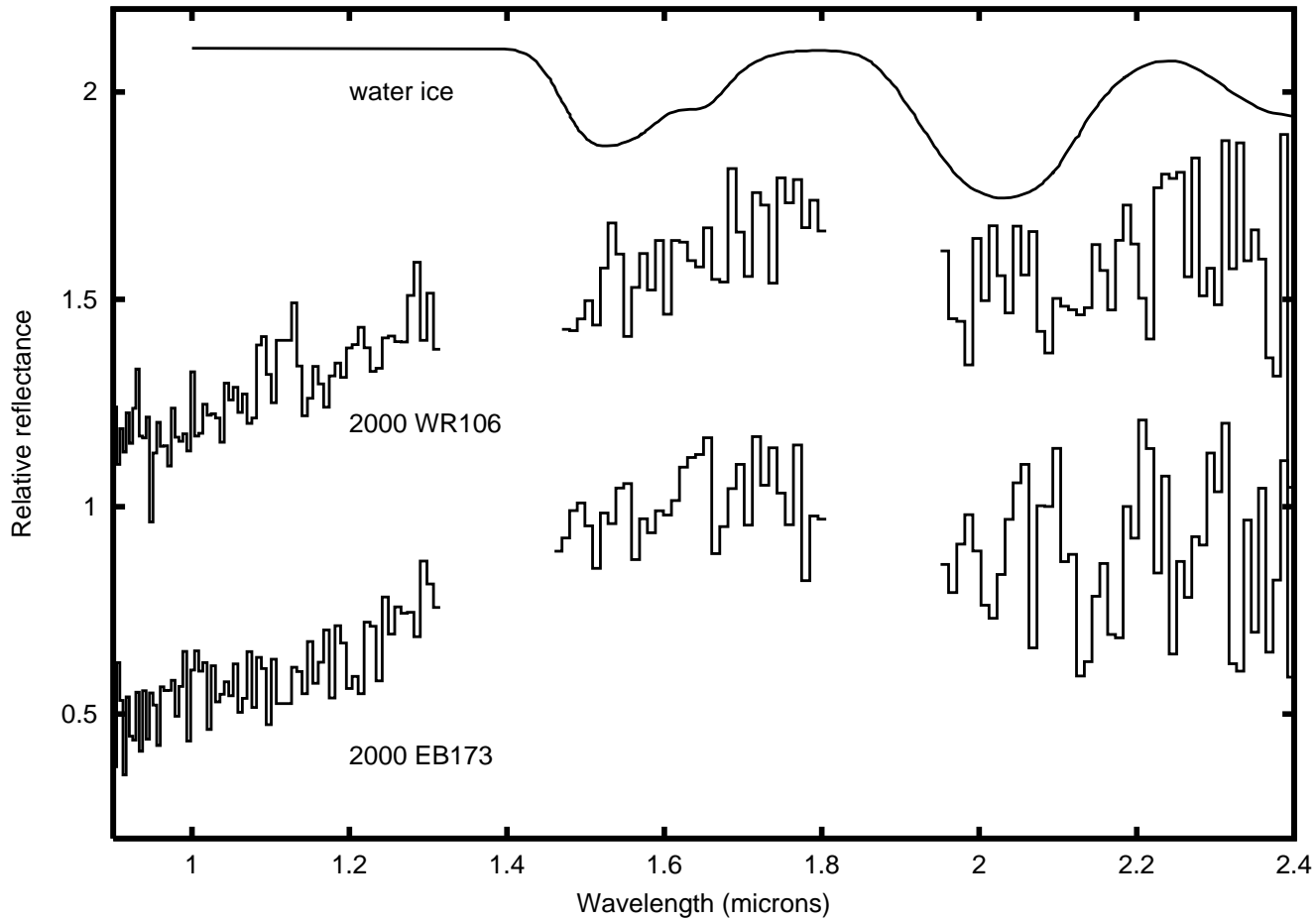}
   \caption{Reflectance spectra of 2000 EB173 and 2000 WR106. The spectra has been
   normalized at 1.7 $\mu$m. 
   The spectrum of 2000 WR106 is shifted. The reflectance spectrum of
   10 $\mu$m water ice grains at 90 K, taken from 
   Brown \& Koresko (\cite{BrownKo98}) is also included. Note the 1.5 and 2.0 
   $\mu$m water ice absorption bands on the spectrum of 2000 WR106.
   }
              \label{Fig1}
    \end{figure*}

   To correct for telluric absorption, A0 stars were  observed during the same night
   (SAO 058110 and SAO 042804 on January 14, and  SAO 058110, SAO 042804, and SAO 018946 on
   February 5). The spectra of the TNOs were divided by the spectra of the A0 stars, multiplied by
   the ratio of the blackbody function of the A0 stars, divided by that of the Sun, and then normalized
   to unity at 1.7 $\mu$m, thus obtaining the relative reflectance plotted in Fig.
   \ref{Fig1}. Around the telluric water band absorptions the S/N of the spectrum is
   very strong, and also this absorption varies between the TNO spectra and the 
   standard stars spectra introducing false spectral features. 
   Therefore, these parts are not included in the final spectrum.
   Finally, the J magnitude of 2000 EB173
   was obtained by means of aperture photometry on the J$_s$ images (J=18.2 +/- 0.11).
   
%   \input{table1}

%                                     Two column figure (place early!)
%______________________________________________ Gamma_1 (lg rho, lg e)
%
%                                     Two column figure (place early!)
%______________________________________________ Gamma_1 (lg rho, lg e)
   \begin{figure*}
   \centering
   \includegraphics[angle=-90,width=\textwidth]{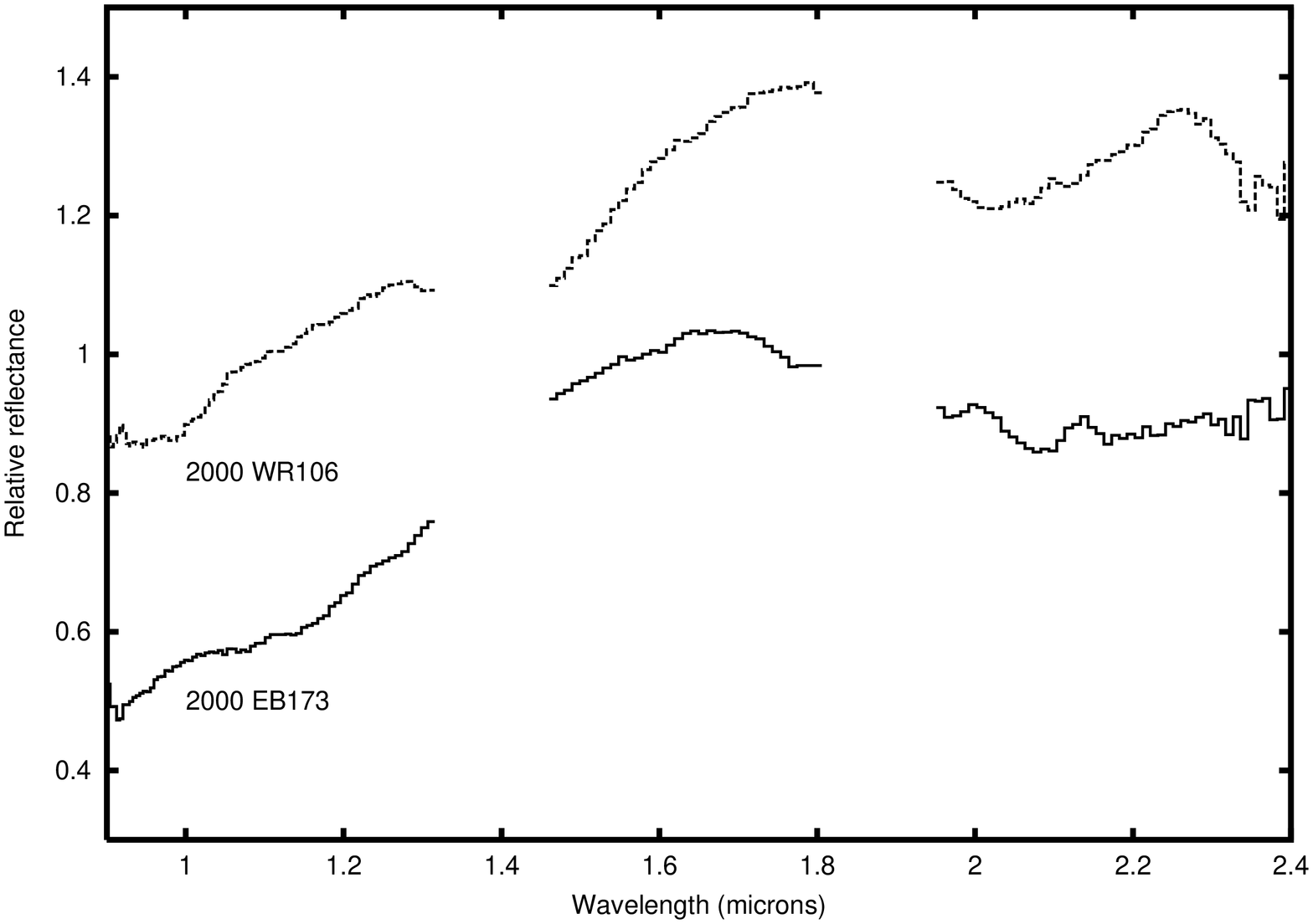}
   \caption{Smoothed version (box-car smoothing of 15 pixels) 
   of the spectra shown in Fig. \ref{Fig1}. Note that both
   spectra present strong absorptions in the K band. 2000 WR106 spectrum also shows 
   signatures of water ice in the surface, at 2.0 and 1.5 $\mu$m.}
              \label{Fig2}
    \end{figure*}
%
 
%__________________________________________________________________

\section{Discussion}

   The spectra of both objects are shown in Fig. \ref{Fig1} together with
   the reflectance spectrum of 10 $\mu$m water ice grains at 90 K, taken from 
   Brown \& Koresko (\cite{BrownKo98}). Both TNOs spectra
   present a very red slope between 0.9 and 1.8 $\mu$m,
   equivalent to J-H = 0.7 and J-H = 0.8 for 2000 WR106 and 2000 EB173, respectively.
   But the slope drastically changes in the K band.
   The turn over between 1.6-1.8 $\mu$m and subsequent depression extending throughout
   the K band is more conveniently visualized in the smoothed spectra displayed
   in Fig. \ref{Fig2}, where one can also notice that the two objects have remarkably
   different spectral behavior within the H and K bands. 

   The spectrum of 2000 WR106 shows absorption features at 1.5 and 2.0 $\mu$m
   typically produced by water ice in the surface. 
   On the other hand, the absorption band
   observed in the K region in the spectrum of 2000 EB173 is much more extended and flat. Also
   no signs of any absorption at 1.5 $\mu$m is detected. We should conclude that water
   ice has not been observed in 2000 EB173 surface. This is in agreement with 
   Brown et al. (\cite{Brownetal2000}), who present a spectrum of 2000 EB173
   in the 1.4 - 2.4 $\mu$m region that is similar to our  in the same bands.
   They also conclude that the spectrum is featureless, but the additional information
   provided by our spectrum in the 0.9 - 1.3 $\mu$m region shows that, 
   in the near-IR spectral
   range, absorption features exist.
      
   The observed very red color of both objects is probably due to an evolved surface 
   which is the result of
   long term irradiation by solar radiation, solar wind, and galactic cosmic-rays. This
   results in the selective loss of hydrogen and the
   formation of an "irradiation mantle" of carbon residues (Moore et al. \cite{Mooreetal83};
   Johnson et al. \cite{Johnetal84}; Strazzulla \& Johnson \cite{StrazzullaJo91}). This process
   makes that an initially neutral color and high albedo ice becomes redish. Further irradiation
   gradually reduces the albedo at all wavelengths, and the material becomes very dark, neutral
   in color, and spectrally featureless (Andronico et al. \cite{Androetal87}; Thompson et al.
   \cite{Thompsonetal87}). The surface water ice present in 2000 WR106 and the fact that
   it is less red than the surface of 2000 EB173 suggest that the thickness of
   2000 WR106 irradiation mantle is smaller than that of 2000 EB173. There are two possible
   explanations for this: the irradiation processes in 2000 EB173 were more intense than
   in 2000 WR106 thus more "fresh" original material remains in the surface; 
   or the surface of 2000 WR106, was more affected by collisions. 
   In the second case, collisions affected deeper layers of the irradiation
   mantle of 2000 WR106 than of 2000 EB173, reaching "fresh" ice under the irradiation mantle
   (Gil-Hutton, personal communication). This fresh material can be also affected by collisions
   (Hartmann \cite{Hartmann}).

   It is also very interesting to note that the turn-over in H and drop in K  
   seems to be usual in the spectra of similar red objects. That is the case of Centaurs
   8405 Asbolus (Barucci et al. \cite{Baruccietal00}; Kern et al. \cite{Kernetal00}),  
   5145 Pholus (Cruikshank et al. \cite{Cruikshanketal98}),
   and TNOs 1996 TP$_{66}$ and 1996 TS$_{66}$ (Noll et al. \cite{Nolletal00}). In some cases, like
   5145 Pholus and 2000 WR106, water ice absorption bands could be in principle considered as 
   the most important contributor to the observed turn over. 
   But is it not the case of 2000 EB173. Also Centaur 8405 Asbolus is a very interesting case, as
   it presents a discrete region composed mostly of water ice superposed on a darker surface according to 
   Kern et al.(\cite{Kernetal00}).
   They observed that the spectrum of this object (1.1 - 1.9 $\mu$m) varies during the 1.7 hours
   of the observations. At the beginning a broad absorption band at 1.6 $\mu$m is clearly detected
   (corresponding to the spot of water ice) and dissapears in the late three spectra, when the 
   water ice spot is not visible. But the change in the slope at about 1.7 $\mu$m remains. 
   Barucci et al. (\cite{Baruccietal00}) also observed this turn over, but not the 
   typical water ice absorption bands, and show that 8405 Asbolus also presents a red slope
   in the 0.4 - 1.7 $\mu$m region. So, in this very red object, even in the absence of water ice,
   the change of the slope of the spectrum in the H band is very pronounced. We should conclude
   that there is a strong and very extended absorption band in the K  region in all observed
   objects that present an irradiation mantle. It should be expected that 
   the materials that produce this absorption probably originate from the same process. 
   Materials like Titan tholin and nitrogen-bearing molecules can contribute 
   significatively (see Cruikshank et al. \cite{Cruikshanketal98}).
   More modeling of the observed spectra and laboratory experiments are needed.

 %  The detection of water ice in 2000 WR106 could also indicate that
 %  the assumption of a low albedo is in error. According to Burns (\cite{Burns86}),
 %  there are solar system objects with the signature of water ice in their spectra that
 %  have visible albedos in the range $p =$ 0.2-0.1. Thus the object could be 
 %  smaller than assumed. In fact Jewitt \& Ausell (\cite{JewittAu01}) derived a geometrical albedo
 %  of 7 \% for this object and give a diameter value of 900 km.

   Finally, low resolution spectroscopy in the whole 0.9 - 2.4 $\mu$m range of a significant
   number of TNOs and Centaurs is crucial for the understanding of their surface properties and
   the physical process that control their evolution. 
         
%   Also similar objects with neutral colors present a large variation in the detected amount of water ice
%   in the surface. Brown et al. (\cite{BrownCrPe99}) detected water ice in 1996 TO$_{66}$ 
%   while Luu \& Jewitt (\cite{LuuJew98}) did not
%   in 1996 TL$_{66}$, though it seems that the diversity between red and neutral TNOs is not
%   due to the presence of water ice in the surface. Thus more infrared spectroscopy of a large number
%   of TNOs is needed to learn the processes that determine the composition of the TNO's surface.
%   For this purpose, an instrument like NICS demonstrate to be very useful. 

\begin{acknowledgements}
   We thank Dale P. Cruikshank for his useful comments on the manuscript and
   Leonardo Testi for his help in the reduction of the spectra.

   This paper is based on observations made
   with the Italian Telescopio Nazionale Galileo (TNG) 
   operated on the island of La Palma by the Centro Galileo
   Galilei of the CNAA (Consorzio Nazionale per l'Astronomia e l'Astrofisica) 
   at the Spanish Observatorio del
   Roque de los Muchachos of the Instituto de Astrofisica de Canarias.
   We are grateful to all the technical staff and telescope operators for
   their assistance during the commissioning phase of NICS.
\end{acknowledgements}

%%%%%%%%%% Bibliografia %%%%%%%%%%%%%%%%%%%%%

\end{document}